\documentclass[aps,preprint, onecolumn,showpacs]{revtex4-1}
\usepackage{graphicx,bm}
\usepackage{epstopdf}
\usepackage{amsmath}
\usepackage{mathrsfs}
\usepackage{color}
\usepackage{subfigure}

\begin{document}

\title{Ginzburg-Landau theory of the bcc-liquid interface kinetic coefficient}
\author{Kuo-An Wu$^1$, Ching-Hao Wang$^1$, Jeffrey J. Hoyt$^2$, and Alain Karma$^3$}

\affiliation{$^1$Department of Physics, National Tsing-Hua University, 30013 Hsinchu, Taiwan\\
$^2$Department of Materials Science and Engineering and Brockhouse Institute for Materials Research,  McMaster University, 1280 Main Street West, Hamilton, Canada L8S 4L7\\
$^3$Physics Department and Center for Interdisciplinary Research on Complex Systems, Northeastern University, Boston, Massachusetts 02115, USA}
\date{\today}


\begin{abstract}
We extend the Ginzburg-Landau (GL) theory of atomically rough bcc-liquid interfaces [Wu {\it et al.}, Phys. Rev. B \textbf{73}, 094101 (2006)] outside of equilibrium. We use this extension to derive an analytical expression for the kinetic coefficient, which is the proportionality constant $\mu(\hat n)$ between the interface velocity along a direction $\hat n$ normal to the interface and the interface undercooling. The kinetic coefficient is expressed as a spatial integral along the normal direction of a sum of gradient square terms corresponding to different nonlinear density wave profiles.  Anisotropy arises naturally from the dependence of those profiles on the angles between the principal reciprocal lattice vectors $\vec K_i$ and $\hat n$.  Values of the kinetic coefficient for the$(100)$, $(110)$ and $(111)$ interfaces are compared quantitatively  to the prediction of linear Mikheev-Chernov (MC) theory [J. Cryst. Growth \textbf{112}, 591 (1991)] and previous molecular dynamics (MD) simulation studies of crystallization kinetics for a classical model of Fe. Additional MD simulations are carried out here to compute the relaxation time of density waves in the liquid in order to make this comparison free of fit parameter. The GL theory predicts a similar expression for $\mu$ as the MC theory but yields a better agreement with MD simulations for both its magnitude and anisotropy due to a fully nonlinear description of density wave profiles across the solid-liquid interface. In particular, the overall magnitude of $\mu$ predicted by GL theory is an order of magnitude larger than predicted by the MC theory. GL theory is also used to derive an inverse relation between $\mu$ and the solid-liquid interfacial free-energy. The general methodology used here to derive an expression for $\mu(\hat n)$ also applies to amplitude equations derived from the phase-field-crystal model, which only differ from GL theory by the choice of cubic and higher order nonlinearities in the free-energy density.
\end{abstract}
\maketitle

\section{Introduction}

A major determinant of the morphology of crystals grown from the melt far from local thermodynamic equilibrium is the solid-liquid interface kinetic coefficient \cite{Book2009,Actareview2000,Hoytetal2003,Actareview2009}. For atomically rough interfaces, this coefficient $\mu(\hat n)$ is the proportionality constant, defined by the linear relation
\begin{equation}
V=\mu(\hat n)\Delta T  ,
\end{equation}
between interface velocity $V$ and interface undercooling $\Delta T=T_m-T$, where $T_m$ is the melting point. The crystallization rate and hence $\mu$ generally depends on the direction $\hat n$ normal to the interface with respect to a fixed set of crystal axes. Both the magnitude and anisotropy of $\mu$ have been shown theoretically, within the framework of solvability theory \cite{Brener1991}, and computationally using both front-tracking \cite{Ihle2000} and phase-field \cite{Bragardetal2002} approaches, 
to have a crucial influence on dendritic solidification at large growth rates. To date, major progress has been achieved in using molecular dynamics (MD) simulations with embedded-atom-method (EAM) interatomic potentials to compute $\mu$ and its anisotropy for various pure metals (e.g. Ni ,Cu, Mg, and Fe) and different crystal structures (e.g. bcc, hcp, and fcc) \cite{Hoytetal1999,Hoytetal2002,Hoytetal2003,Sunetal2004,Gaoetal2010,Monketal2010}. Moreover, results of those simulations, such as for pure Ni \cite{Hoytetal2002}, have been used as input parameters in phase field simulations \cite{Bragardetal2002}, thereby making it possible to link quantitatively atomistic and continuum length scales for the prediction of dendrite growth rates that have been compared with experiments \cite{Hoytetal2003,Actareview2009}. Furthermore, results from MD simulations have made it possible to test quantitatively basic theories of crystal growth kinetics, thereby shedding light on the physical mechanisms that controls $\mu(\hat n)$ (see \cite{Actareview2009} for a review). The magnitude of $\mu$ has been found to be well predicted by the expression 
\begin{equation}
\mu \approx C\frac{V_TL}{k_BT_m^2}, \label{BGJ}
\end{equation}
proposed by Broughton, Gilmer, and Jackson (BGJ) to interpret crystallization rates measured by MD simulations in the Lennard-Jones system \cite{BGJ1982}. Here
$V_T=\sqrt{3 k_BT/m}$ is the thermal velocity of atoms in the liquid, assumed to limit the rate of atomic attachment at the interface, $m$ is the atomic mass and $C$ is a constant of order unity that can generally depend on the growth orientation; $L$ is the latent heat per atom. BGJ introduced Eq. (\ref{BGJ})  based on the finding that crystallization rates were too large to be explained by the common assumption that atomic attachment at the solid-liquid interface is a thermally activated process with the same energy barrier as liquid-state diffusion. Coriell and Turnbull \cite{Coriell1982} independently developed an expression for $\mu$ in metallic systems based on a similar assumption that crystallization is limited by the rate of liquid-atom collisions at the interface, but related this rate 
to the frequency of atomic vibrations in the solid instead of to the thermal velocity of liquid-atoms, which yields the expression
$\mu\approx V_SL/k_BT_m^2$ where $V_S$ is the speed of sound in the solid. This upper bound estimate of $\mu$ is much larger than values extracted from MD simulations to date for pure metals \cite{Hoytetal1999,Hoytetal2002,Hoytetal2003,Sunetal2004,Gaoetal2010,Monketal2010}, which are in closer agreement with Eq. (\ref{BGJ}).  
 
Eq. (\ref{BGJ}) has been put on a firmer theoretical footing by 
Mikheev and Chernov (MC) \cite{MC1991,Chernov2004} who derived a similar form in the theoretical framework of classical density functional theory of freezing \cite{DFT1,DFT2,DFT3}. In this density wave picture, crystal ordering of atoms increases from liquid to solid through several atomic layers parallel to the interface. Concomitantly, the amplitude of density waves corresponding to different reciprocal lattice vectors (RLV) of the crystal lattice increases smoothly from zero in the liquid to constant values in the crystal along the $z$-axis normal to the interface. The expression for $\mu$ in the MC theory is derived by only considering the contribution of the set of principal RLV (with lowest magnitude $|\vec K_i|$) to the crystal density field, and by using a fluctuation-dissipation relation to relate the rate of interface dissipation to the inverse half-width of the dynamic equilibrium structure factor  $S(|\vec K_i|,\omega)$ \cite{cmt}. This theory predicts a magnitude of $\mu$ of the form of Eq. (\ref{BGJ}) where $C$ depends on growth orientation through the orientation dependence of the spatial decay rate of density waves into the liquid, which depends on $\hat{K}_i\cdot \hat n$. 
It should also be noted that, according to the MC theory, the kinetic coefficient depends on a relaxation time of density waves in the liquid, which in turn can be related to the liquid diffusion coefficient.  Therefore, the MC model appears to disagree with the main assumption first proposed by BGJ.  In a recent MD study,
Mendelev \emph{et al.} \cite{Mendelev} showed that, at least in the limit of small undercoolings, $\mu$ is in fact proportional to the diffusivity.  The authors speculate that there is a change in atomic attachment mechanism in the high and low undercooling limits.
 
In this paper, we derive an expression for $\mu$ within the framework of Ginzburg-Landau (GL) theory. Like the MC theory, GL theory is rooted in a density wave picture of the solid-liquid interface structure and considers a minimal set of RLV to represent the crystal density field. However a non-trivial difference between the two theories is that the MC theory is linearized around the liquid state and hence neglects the nonlinear interaction between different density waves. The anisotropy of $\mu$ arises solely in this theory from the orientation dependence of the exponential decay rate into the liquid of non-interacting density waves. In contrast, GL theory captures the nonlinear interaction between different density waves through the inclusion of cubic and quartic terms in the GL expansion of the free-energy density in density wave amplitudes. Consequently, the resulting expression for $\mu$ derived here in the GL framework depends explicitly on the interacting nonlinear density wave profiles through the entire solid-liquid interface region and, as a result, $\mu$ has a different anisotropy than that predicted by the linearized MC theory \cite{MC1991,Chernov2004}. 

We carry out our analysis for the bcc-liquid interface whose equilibrium properties, in particular the excess free-energy of the interface $\gamma_{sl}$ and its anisotropy,  have been modeled previously by GL theory \cite{shih,bcc1}. This equilibrium theory is extended to a non-equilibrium situation in the standard framework of the time-dependent GL (TDGL) theory. We incorporate a thermodynamic driving force proportional to the undercooling and a free-energy dissipation time scale that is related, as in the MC theory, to the inverse half-width of the dynamic equilibrium structure factor.  
The kinetic coefficient $\mu$ is calculated explicitly for $(100)$, $(110)$ and $(111)$ interfaces using parameters obtained from MD simulations for the Fe EAM potential developed by Mendelev \emph{et al.} \cite{MHSA2} and the results are compared to the predictions of MD simulations using this potential \cite{Sunetal2004,Gaoetal2010} and the MC theory.  

We note that the general methodology developed here to derive an expression for the kinetic coefficient within a TDGL framework applies directly to amplitude equations for elemental systems \cite{Goletal05,Athetal06,Athetal07,Wu07,Goletal09} and binary alloys \cite{Eldetal2010,SpaKar2010} derived from the phase-field-crystal (PFC) model \cite{Eldetal02,Eldetal04,Eldetal07,Proetal07}. As shown previously by Wu and Karma \cite{Wu07} in a study of the equilibrium bcc-liquid interface, the set of amplitude equations derived from the PFC model only differs from the set derived from GL theory in the coefficients of nonlinear terms that couple different density waves. In the amplitude equations derived from the PFC model, all coefficient of nonlinear terms are uniquely determined by the nonlinear form assumed for the free-energy density in the PFC model from which the amplitude equations are derived. In contrast, in the versions of GL theory of Refs. \cite{shih, bcc1}, those coefficients are determined by the ansatz that all {\it geometrically distinct} closed polygons with the same number of sides corresponding to RLV have equal weight. In principle, the weight of closed polygons in reciprocal space can be derived if higher order $n$-point correlation functions are provided. However, this information is difficult to obtain. If one assumes that higher order correlation functions are constant, then one recovers the nonlinear coefficients in the amplitude equations derived from the PFC model \cite{Wu07,Toth}. Differences in coefficients obtained from an amplitude expansion of the standard PFC model and this ansatz were found to have only a small effect on the prediction of $\gamma_{sl}$ and its anisotropy for the bcc-liquid interface \cite{Wu07}. However, more generally, the formalism developed in the present work should prove useful in the development of PFC formulations and amplitude equations that model different kinetic anisotropies for different crystal structures. 

We first write down the TDGL model of crystallization and then use this model to derive an analytical expression for the kinetic coefficient. We detail the procedure for a specific choice of orientation and state the results for other orientations. Next, we present a method to compute the relaxation time of density waves in the liquid that is a key kinetic input parameter for both the MC and GL theories. We then compare the predictions of GL theory to the predictions of the linearized MC theory and previous MD simulation studies. 

\section{Time-dependent Ginzburg-Landau model}

To construct a TDGL model of crystallization kinetics for the bcc-liquid system, we start from the expression    
for the excess free-energy $\Delta F$ for the solid-liquid system in equilibrium relative to the liquid free energy. Under the assumption that the density wave amplitude varies slowly on the scale of the lattice spacing, this excess has the form \cite{bcc1}: 
\begin{eqnarray}\label{equF}
\Delta F&\approx&\frac{n_0k_BT}{2}\int d\vec{r}\left(\sum_{i,j}\frac{1}{S(|\vec{K}_i|)}u_iu_j\delta_{0,\vec{K}_i+\vec{K}_j}\right.\nonumber\\
&+&b\sum_i \left. c_i  \left|\frac{du_i}{dz}\right|^2\right.\nonumber\\
&-&\left.a_3\sum_{i,j,k}c_{ijk}u_iu_ju_k\delta_{0,\vec{K}_i+\vec{K}_j+\vec{K}_k}\right.\nonumber\\
&+&\left.a_4\sum_{i,j,k,l}c_{ijkl}u_iu_ju_ku_l\delta_{0,\vec{K}_i+\vec{K}_j+\vec{K}_k+\vec{K}_l}\right),
\end{eqnarray}
The $u_i$'s denote the amplitudes of density waves corresponding to the RLV with the smallest magnitude $|\vec K_i|$ in the truncated expansion of the number density 
\begin{equation}\label{density}
n(\vec{r},t)=n_0 \left( 1+ \sum_{\vec{K}_i}u_{i}(\vec{r},t)e^{i\vec{K}_i\cdot \vec{r}}+\dots \right), 
\end{equation}
and have the limits $u_i=u_s$ ($u_i=0$) in the solid (liquid). Since the reciprocal lattice of bcc is fcc, there are twelve $|\vec K_i|$'s of equal magnitude pointing in $\langle 110\rangle$ directions.
$S(K)$ denotes the liquid structure factor and
$C(K)$ refers to the Fourier transform of the direct correlation function $C(|\vec{r}-\vec{r}'|)$ and $C''(K)\equiv d^2C(K)/dK^2$. 
The coefficients of the gradient square terms are determined by comparison of the form \eqref{equF} and the expression for the free-energy of an inhomogeneous liquid, yielding $b=-2C''(|\vec K_i|)$ and $c_i=(\hat{K}_i\cdot\hat{n})^2/4$ \cite{bcc1}.
The coefficients $a_3$ and $a_4$ are determined in the same way as in Shih \emph{et al.} \cite{shih} and Wu \emph{et al.} \cite{bcc1} from the 
the two equilibrium conditions that the solid and liquid phases must have equal free energies
at the melting point and the equilibrium state of the solid is a minimum of free-energy. These two conditions yield
the values $a_3=2a_2/u_s$ and $a_4=a_2/u_s^2$ where $a_2=12/S(|\vec K_i|)$. In addition, the aforementioned ansatz that all closed polygons of $\vec K_i$'s with the same number of sides have equal weight yields the constants $c_{ijk}=1/8$ and $c_{ijkl}=1/27$. 

It is important to note that in the absence of knowledge of higher order correlation functions there is no general way to determine the weight for each closed polygon. Thus we simply assume that geometrically distinct polygons (i.e., exclude repetitive polygons) have equal weight. However, if one assumes that Fourier transforms of higher order correlation functions are constant, then all polygons (including repetitive polygons) contribute equally in the free energy and this yields $c_{ijk}=1/48$ and $c_{ijkl}=1/540$ as shown in the PFC calculations \cite{Wu07, Toth}. It is straightforward to examine the relation between these normalization constants. For example, there are 27 geometrically distinct 4-side polygons. Out of these 27 polygons, 6 of them contain twice the same RLVs (e.g., $[110]$, $[\bar{1} \bar{1} 0]$, $[110]$, $[\bar{1} \bar{1} 0]$), and 21 of them contain 4 different RLVs (e.g., $[110]$, $[1 \bar{1} 0]$, $[\bar{1} 10]$, $[\bar{1} \bar{1} 0]$). Thus if we choose to count all repetitive polygons, the number of 4-side polygons is $4!/(2! 2!)\times 6 + 4!\times 21 = 540$ (since there are $4!/(2!2!)$ ways to rearrange RLVs for the 6 polygons that contain twice the same RLVs and $4!$ ways for each of the 21 polygons that contain four different RLVs).

To incorporate a driving force for crystallization in the model, we expand the free energy difference between the solid and liquid phases near the melting point in the form
\begin{equation}\label{Fadd}
F_S(T)-F_L(T)=(S_S-S_L)(T-T_m)=L\frac{T-T_m}{T_m},
\end{equation} 
where we have used the thermodynamic relation $dF=-SdT$ and $L$ denotes the latent heat of melting per atom.  Furthermore, we add this driving force by assuming that this free-energy difference varies proportionally to the density wave amplitude through the solid-liquid interface region. This yields the expression for the free-energy of the two-phase system outside of equilibrium 
\begin{equation}
\Delta F'=\Delta F+{n_0k_BT_m}\int d\vec{r}\sum_i\frac{1}{12}\frac{u_i-u_s}{u_s}\frac{L}{k_BT_m}\frac{T-T_m}{T_m}.
\end{equation}  
The normalization constant $1/12$ in the driving force term ensures that for bcc lattices the bulk energy difference between solid and liquid has the correct temperature dependence imposed by Eq. \eqref{Fadd}.

Next, we assume that the evolution of the order parameters $u_i$ is governed by an equation of the standard TDGL form
\begin{equation}\label{EOM0}
\tau\frac{\partial u_i}{\partial t}=-\frac{1}{n_0k_BT}\frac{\delta \Delta F'}{\delta u_i},
\end{equation}
where the kinetic time scale $\tau$ is fixed by the requirement that density waves in the liquid should relax on a time scale $\tau_L(|\vec K_i|)$ corresponding to the inverse half-width of the dynamical structure factor $S(|\vec K_i|,\omega)$. This requirement is satisfied by the choice 
\begin{equation}
\tau=\tau_L(|\vec K_i|)/S(|\vec K_i|).
\end{equation}
With the above choice, the TDGL equation (\ref{EOM0}) reduces in the liquid to $\tau_L(|\vec K_i|) \partial_t u_i = - u_i$ due to the cancellation of the factor of $1/S(|\vec K_i|)$ on both sides of the equation.

\section{Analytical calculation of the kinetic coefficient}
 
To derive an expression for the kinetic coefficient, we look for a steady-state propagating solution of the TDGL equation that corresponds to planar crystallization fronts moving at constant velocity $V$. Those solutions have the general form
$u_i(\vec r,t)=u_i(\hat n\cdot \vec r-Vt)$ where $\hat n$ is the crystal growth direction normal to the solid-liquid interface. To analyze those  solutions, we transform Eq. \eqref{EOM0} to a moving frame translating at velocity $V$ along the normal direction through the coordinate transformation $z=\hat n\cdot \vec r-Vt$, which yields the set of coupled nonlinear ordinary differential equations
\begin{equation}\label{EOM}
-V\tau\frac{d u_i}{d z}=-\frac{1}{n_0k_BT}\frac{\delta \Delta F'}{\delta u_i},
\end{equation}
for the time-independent profiles $u_i(z)$. 
For a given direction of the interface $\hat{n}$, an analytic expression for $\mu$ can be obtained by looking for solutions of Eq. \eqref{EOM} in the limit of small driving force where $V\sim \Delta T$ and the propagating density wave profiles deviate only slightly from the stationary equilibrium profiles for $V=0$. In this limit, the problem of finding solutions to Eq. \eqref{EOM} can be transformed into a linear problem by linearizing Eq. \eqref{EOM} around the equilibrium 
profiles, i.e. by substituting $u_i(z)=u_{i0}(z)+u_{i1}(z)+\dots$ where $u_{i0}(z)$ denote the stationary equilibrium profiles and $u_{i1}(z)$ denote small linear perturbations of those equilibrium profiles due to interface motion. An expression for $\mu$ is then readily obtained from the solvability condition  of finding the solutions $u_{i1}(z)$ to a set of coupled linear differential equations with some non-constant coefficients that depend on the $u_{i0}(z)$ profiles. This procedure is a straightforward generalization of the standard procedure used to derive an expression for the interface kinetic coefficient in the standard single order parameter phase-field model of crystal growth (e.g., see \cite{KarRap1998}). We carry out this calculation explicitly below for the three low index crystal faces generally considered in characterizing the anisotropy of interface properties in fcc- and bcc-forming systems.  

To start, we use the results of previous work on capillary anisotropy for bcc-liquid interfaces \cite{bcc1}. This analysis shows that the amplitudes of density waves can be categorized into different groups according to the relative orientations of different principal RLV. Those orientations determine the values of $\hat{K}_i\cdot \hat n$ and hence the coefficients of the square gradient terms appearing in the GL free-energy functional \eqref{equF} as summarized in Table \ref{cis} for the three crystal faces considered. To exemplify our calculation in detail, we choose the $(110)$ crystal face for which the amplitude of propagating density waves are denoted as $u$, $v,$ and $w$ with corresponding values of $(\hat{K}_i\cdot \hat{n})^2$, $1/4$, $1$, and $0$, respectively.

 \begin{table}[b]
 \caption{
Classifications and the values of square gradient term $c_i$ for different orientations of bcc crystal interfaces.}
\centering
\begin{tabular*}{1.0\textwidth}%
     {@{\extracolsep{\fill}}c||cc|ccc|cc}  \hline
 & \multicolumn{2}{c}{$100$}&\multicolumn{3}{c}{$110$}&\multicolumn{2}{c}{$111$}
\\ \hline
$(\hat{K}_{i} \cdot \hat{n})^2$ & 0 &1/2& 1/4 & 1& 0 &0&2/3  \\ \hline
Number of $\vec{K}_{i}$'s    & 4 & 8 & 8 &2 &2 &6& 6 \\ \hline
$c_i=(\hat{K}_i\cdot\hat{n})^2/4$   & 0 &1/8&  1/16&1/4& 0 & 0&1/6    \\ \hline
\end{tabular*}
\label{cis}
\end{table}

\begin{table*}[h]
\caption{Values of input parameters from MD simulations with interatomic EAM potential for Fe from MH(SA)$^2$ \cite{MHSA2, Sunetal2004} and resulting coefficients used in GL theory. The value of $\tau_L$ is computed using the method described in section \ref{sec:tau}.} 
\centering
\begin{tabular*}{1.0\textwidth}%
     {@{\extracolsep{\fill}}cccccccccc} \hline
 &$n_0$ ($\AA^{-3}$)&  $a_2$ &$b$ ($\AA^2$) &$\tau_L$ (ps) & $u_s$ & $|\vec{K}_{i}|$ ($\AA^{-1}$) & $\xi_b$ ($\AA$)& $L$ (eV/atom) & $T_m$ (K) \\ \hline
MD [MH(SA)$^2$] & 0.0765&3.99 & 20.81 & $0.57 \pm 0.05$ &0.72 & 2.985 & 3.96 & 0.162 & 1772  \\ \hline
\end{tabular*}
\label{tabparam}
\end{table*}

We write down explicitly Eq. \eqref{EOM} for the three order parameters $u$, $v$ and $w$   
\begin{eqnarray}\label{EOM3}
-4V\tau\frac{d u}{d z}&=&-\left(\frac{1}{2}f_u +2\, C''(|\vec{K}_{110}|)(\hat{K}_u\cdot\hat{n})^2\frac{d^2u}{dz^2} + 4\alpha\right)\nonumber\\
-V\tau\frac{d v}{d z}&=&-\left(\frac{1}{2}f_v+\frac{1}{2}\, C''(|\vec{K}_{110}|)(\hat{K}_v\cdot\hat{n})^2\frac{d^2v}{dz^2}+\alpha\right)\\
-V\tau\frac{d w}{d z}&=&-\left(\frac{1}{2}f_w+ \frac{1}{2}\, C''(|\vec{K}_{110}|)(\hat{K}_w\cdot\hat{n})^2\frac{d^2w}{dz^2}+\alpha\right),\nonumber
\end{eqnarray}
where we have defined the dimensionless parameter 
\begin{equation}
\alpha=\frac{L (T-T_m)}{12 u_s k_BT_m^2 }
\end{equation}
that measures the departure from equilibrium
and used the shorthand notation of partial derivatives of the bulk free-energy density $f$ at equilibrium (defined as $\Delta F= n_0 k_B T \int d\vec{r}f(u,v,w)$) with respect to the order parameters $f_u\equiv\partial f/\partial u$, $f_v\equiv\partial f/\partial v$, and $f_w\equiv\partial f/\partial w$.
As outlined earlier,  we now expand the moving profiles for a temperature slightly below the melting point around the equilibrium profiles at the melting point in the form $u=u_0+u_1+\dots$, $v=v_0+v_1+\dots$, and $w=w_0+w_1+\dots$ where $u_0, v_0$ and $w_0$ denote the equilibrium profiles that are solutions of Eq. \eqref{EOM3} for $V=\alpha=0$ and $u_1, v_1$ and $w_1$ denote the perturbation of those profiles due to interface motion below the melting point. Linearizing Eq. \eqref{EOM3} around the stationary equilibrium profiles, we obtain a set of coupled linear equations for  $u_1, v_1$ and $w_1$. It is convenient to write those linearized equations in the matrix notation
\begin{equation}\label{matrixeqn}
LU=F,
\end{equation}
where we have defined  
\begin{equation}\label{matrixL}
L=  \begin{pmatrix}
                   f_{uu}+4\mathcal{D}_u&f_{uv}&f_{uw}\\
                   f_{vu}&f_{vv}+\mathcal{D}_v&f_{vw}\\
                   f_{wu}&f_{wv}&f_{ww}+\mathcal{D}_w
            \end{pmatrix},
\end{equation}
and 
\begin{eqnarray}
\mathcal{D}_i&\equiv& C''(|\vec{K}_{110}|)(\hat{K}_i\cdot\hat{n})^2\frac{d^2}{dz^2}\\
U&=&  \begin{pmatrix}
                   u_1\\
                   v_1\\
                   w_1
            \end{pmatrix},\quad
F=  2 \begin{pmatrix}
                   4V\tau\frac{d u_0}{d z}-4\alpha\\
                   V\tau\frac{d v_0}{d z}-\alpha\\
                   V\tau\frac{dw_0}{dz}-\alpha,
                               \end{pmatrix}.
\end{eqnarray}
A solvability condition for the existence of a solution to this inhomogeneous linear problem can be readily obtained by noting two properties of the linear operator. First, owing to the translational invariance of the TDGL equation, the right column vector function $U_0$ with components $\frac{du_0}{dz}$, $\frac{dv_0}{dz}$, and $\frac{dw_0}{dz}$ is a solution of the homogeneous linear problem $LU_0=0$, which can be seen explicitly by differentiating Eq. \eqref{EOM3} at the melting point ($V=\alpha=0$) with respect to $z$. Second, the operator $L$ is self-adjoint so that left zero-modes are identical to right zero-modes. This implies that, for any $U$,  $L$ satisfies the property
$(U_0^T,LU)=(U^T,LU_0)=0$ where $U_0^T$ is the transposed left row vector function $U_0^T=(\frac{du_0}{dz},\frac{dv_0}{dz},\frac{dw_0}{dz})$ and $(g,h)=\int_{-\infty}^{+\infty}dz \,  g\cdot h$ denotes the inner product of a left row vector function $g$ and a right column vector function $h$. The first equality $(U_0^T,LU)=(U^T,LU_0)$ can be easily verified using the fact that $L$ is a symmetric matrix and integrating by parts twice over $z$ the diagonal second derivative terms; boundary terms vanish owing to the property that spatial derivatives of $u_0$, $v_0$ and $w_0$ vanish at $z=\pm \infty$. The second equality $(U^T,LU_0)=0$ follows from the first property $LU_0=0$. Hence, for Eq. \eqref{matrixeqn} to have a non-trivial solution, 
we must have $(U_0^T,F)=(U_0^T,LU)=(U^T,LU_0)=0$, yielding
the solvability condition
\begin{eqnarray}
(U_0^T,F)&=&\int_{-\infty}^\infty dz \,2V\tau\left\{\left[4\left(\frac{d u_0}{dz}\right)^2+\left(\frac{d v_0}{dz}\right)^2+\left(\frac{d w_0}{dz}\right)^2\right]\right. \nonumber\\ 
&-&\left.2 \alpha\left[4\frac{d u_0}{dz}+\frac{d v_0}{dz}+\frac{d w_0}{dz}\right]\right\} 
=0. 
\end{eqnarray}  
Setting the boundary conditions for a solid-liquid system $u_0(-\infty)=v_0(-\infty)=w_0(-\infty)=0$ and $u_0(\infty)=v_0(\infty)=w_0(\infty)=u_s$, the density wave velocity $V$ can be further simplified into (here the subscript of V indicates the crystal face normal $(110)$ specific to this case)
\begin{equation}
V_{110}=\frac{12\alpha u_s}{\tau}\left[\int_{-\infty}^\infty dz\,8\left(\frac{d u_0}{dz}\right)^2+2\left(\frac{d v_0}{dz}\right)^2+2\left(\frac{d w_0}{dz}\right)^2\right]^{-1}.
\end{equation}
The growth velocity for other crystal orientations can be computed using the same analysis with references to different sets of density wave amplitudes and square gradient terms listed in Table \ref{cis}. It is clear that the kinetic anisotropy of the solid-liquid interface is a result of different density waves profiles for different crystal orientations. The kinetic coefficient $\mu$ is obtained accordingly by dividing the growth velocity by the undercooling, 
\begin{eqnarray}\label{finalresult}
\mu&=&\frac{12\alpha u_s}{\tau (T-T_m)}\left[\int dz\sum_{\vec{K}_i}\left(\frac{d u_i}{dz}\right)^2\right]^{-1}\nonumber\\
&=&\frac{L}{k_BT^2_m}\frac{S(\vec{K}_i)}{\tau_L(\vec{K}_i)}\left[\int dz\sum_{\vec{K}_i}\left(\frac{d u_i}{dz}\right)^2\right]^{-1}.
\end{eqnarray}
%
%
\section{Computation of the liquid relaxation time from molecular dynamics simulations}
\label{sec:tau}
In order to quantitatively compare the GL model with the results from MD simulation the relaxation parameter $\tau_L ( | \vec K_i | )$ must be determined for the Fe MH(SA)$^2$ potential.  In principle an MD simulation can be performed to determine the dynamic structure factor and, as discussed above, the relaxation time can be found from the inverse half-width of $S( |\vec K_i |, \omega)$. However, we have utilized an alternative method that provides a more convenient and more direct computation of $\tau_L (| \vec K_i |)$.  The MD procedures are as follows.

An 8000 atom simulation cell was melted and subsequently equilibrated for 100 ps at the melting temperature of MH(SA)$^2$ Fe.  During the equilibration the $x$ dimension was held fixed whereas the other two cell dimensions were allowed to vary, such that the pressure in the system was maintained at zero.  The equilibrated liquid was further equilibrated in an NVT ensemble where, in addition to the usual interatomic forces, an force of the form $f = a \cos (|\vec K_i| x)$ was imposed.  Application of the external force results in a one-dimensional number density profile in the liquid with the desired wavenumber $|\vec K_i|$ and the simulation cell length along the $x$ direction, $L_x$, was chosen such that a total of 36 number density peaks are commensurate with the cell dimension (i.e. $L_x = 36 (2 \pi)/|\vec K_i|$).  The optimal choice of the force amplitude $a$ results in a number density amplitude that is sufficiently high to be resolved above the usual thermal fluctuations in density, yet small enough such that the density profile can be accurately described by the form $A(t) \cos (|\vec K_i| x) + n_o$.  By trial and error we found that a value of 0.06 $eV/\AA$ was ideal.  The final step of the $\tau_L$ computation is a short (2 ps) simulation in an NVT ensemble where the external potential is removed.  The exponential decay of $A(t)$ yields directly the relaxation time. In the final simulations a standard Nose-Hoover thermostat was employed and a range of thermostat relaxation parameters from 0.1-1.0 ps were tested.  It was found that the results were unchanged for thermostat settings above 0.5 ps. 

\begin{figure}
\centering
	\includegraphics[width=4.0in]{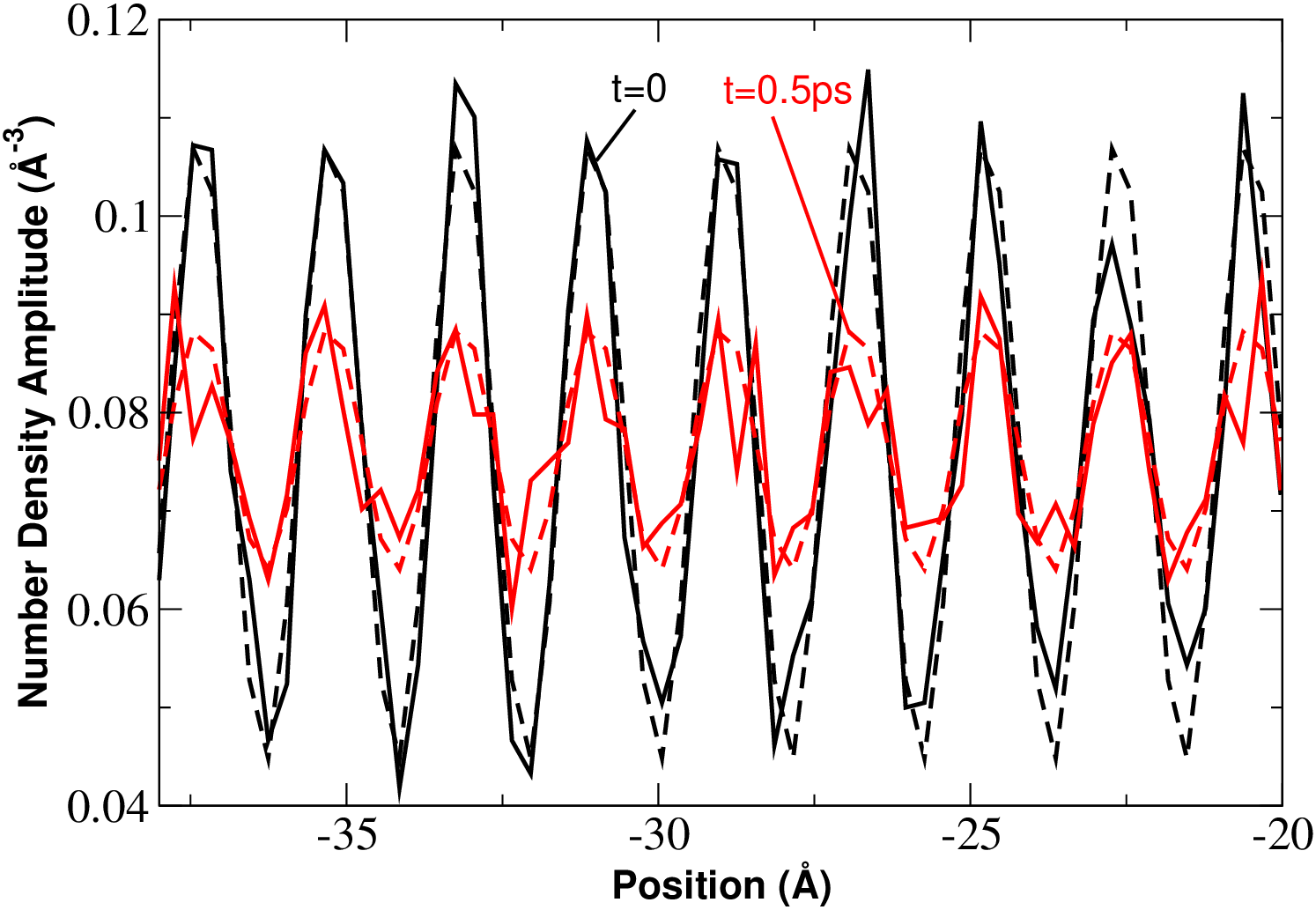} 
\caption{(Color online) The number density in the liquid plotted vs the position $x$ for a portion of the MD simulation cell.  Solid lines are the number densities obtained from the simulation and the dashed lines are best fits to the function $A(t)\cos (|\vec K_i| x) + n_o$. The difference from the initial profile ($t=0$) and a later snapshot ($t=0.5$ ps) illustrates the decay of $A(t)$ with time. }
\label{dens_profile}
\end{figure}

\begin{figure}
\centering
	\includegraphics[width=4.0in]{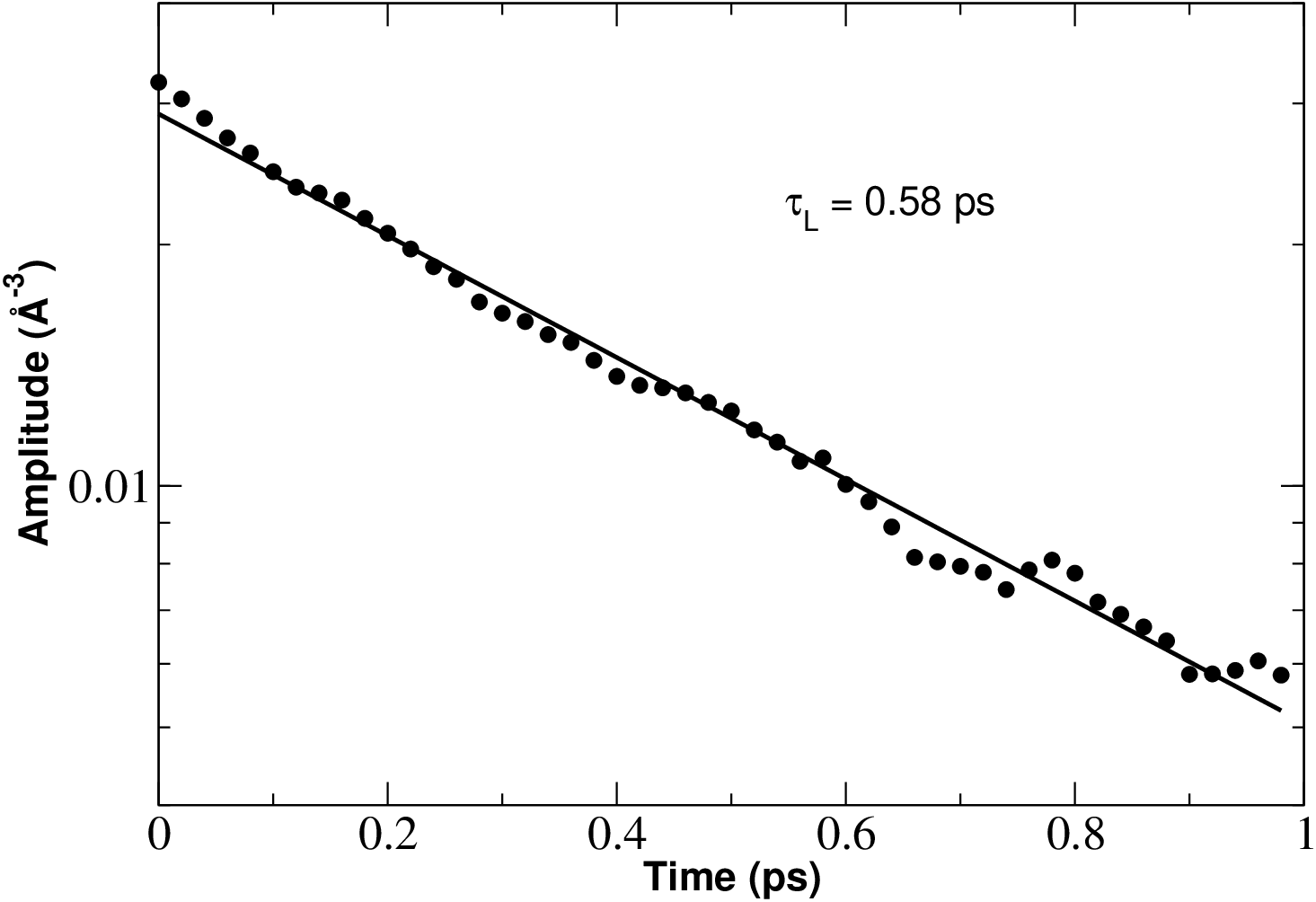} 
\caption{(Color online) Semi-log plot of the amplitude $A(t)$ vs. time for a typical MD simulation.}
\label{ampvst}
\end{figure}

Fig. ~\ref{dens_profile} shows the number density profile at two different times during the decay process.  For clarity, only a portion of the simulation cell is plotted along the $x$ direction and the number density represents the average of five separate runs using different starting configurations.  The high amplitude profile corresponds to the initial profile established in the liquid due to the imposed external force and the dashed line shows the best fit to a cosine function.  The lower amplitude curve corresponds to a time of 0.5 ps and the decay in amplitude is clearly evident.  Fig. ~\ref{ampvst}, plotted on a semi-log scale, illustrates the decay of the best fit amplitude vs time.  The data is well represented by an exponential decay and for this simulation a relaxation time of $\tau_L = 0.58$ ps was found.  In order to assess the statistical uncertainty the above procedure was repeated six times and each computation utilized five different starting configurations for the liquid under an imposed external force.  The final value of the relaxation time was found to be $\tau_L = 0.57  \pm 0.05 $ ps where the error denotes a 95\% confidence limit.


\section{Results and discussion}

\begin{figure}
\centering
\subfigure[]{
	\includegraphics[width=3.0in]{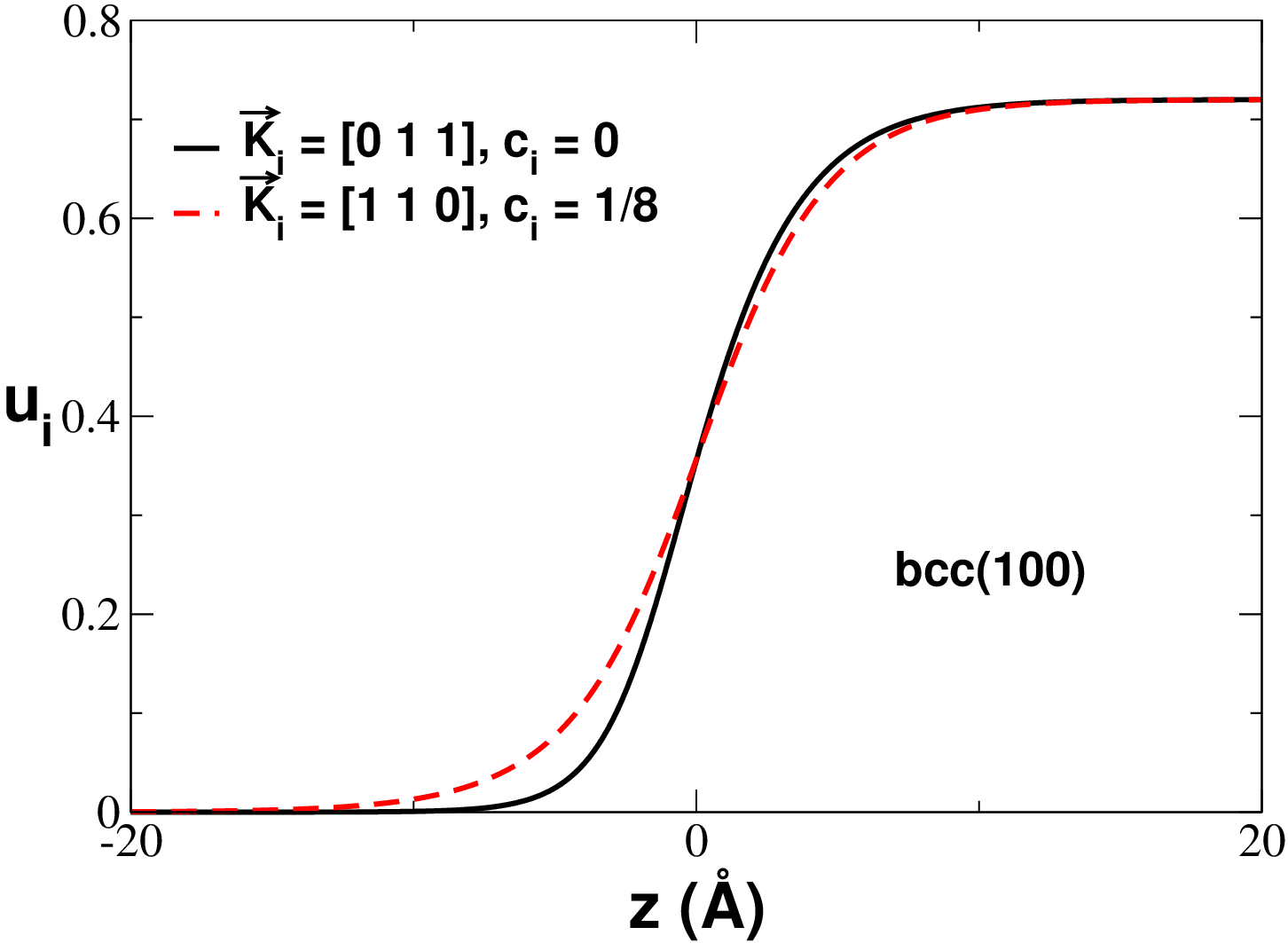} 
	\label{fig:1a}
	} \bigskip \bigskip
\subfigure[]{
	\includegraphics[width=3.0in]{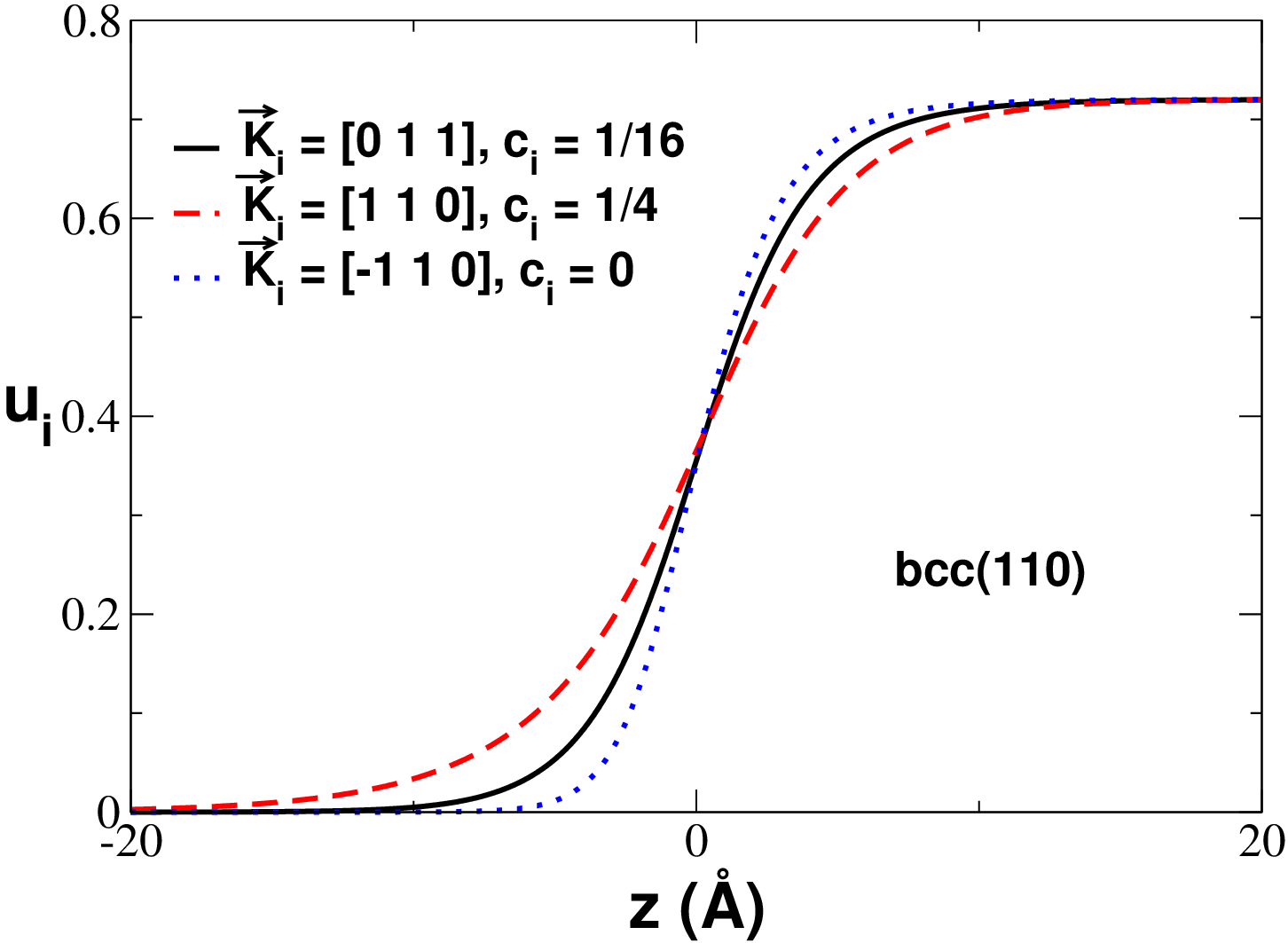} 
	\label{fig:1b}
	} \bigskip \bigskip
\subfigure[]{
	\includegraphics[width=3.0in]{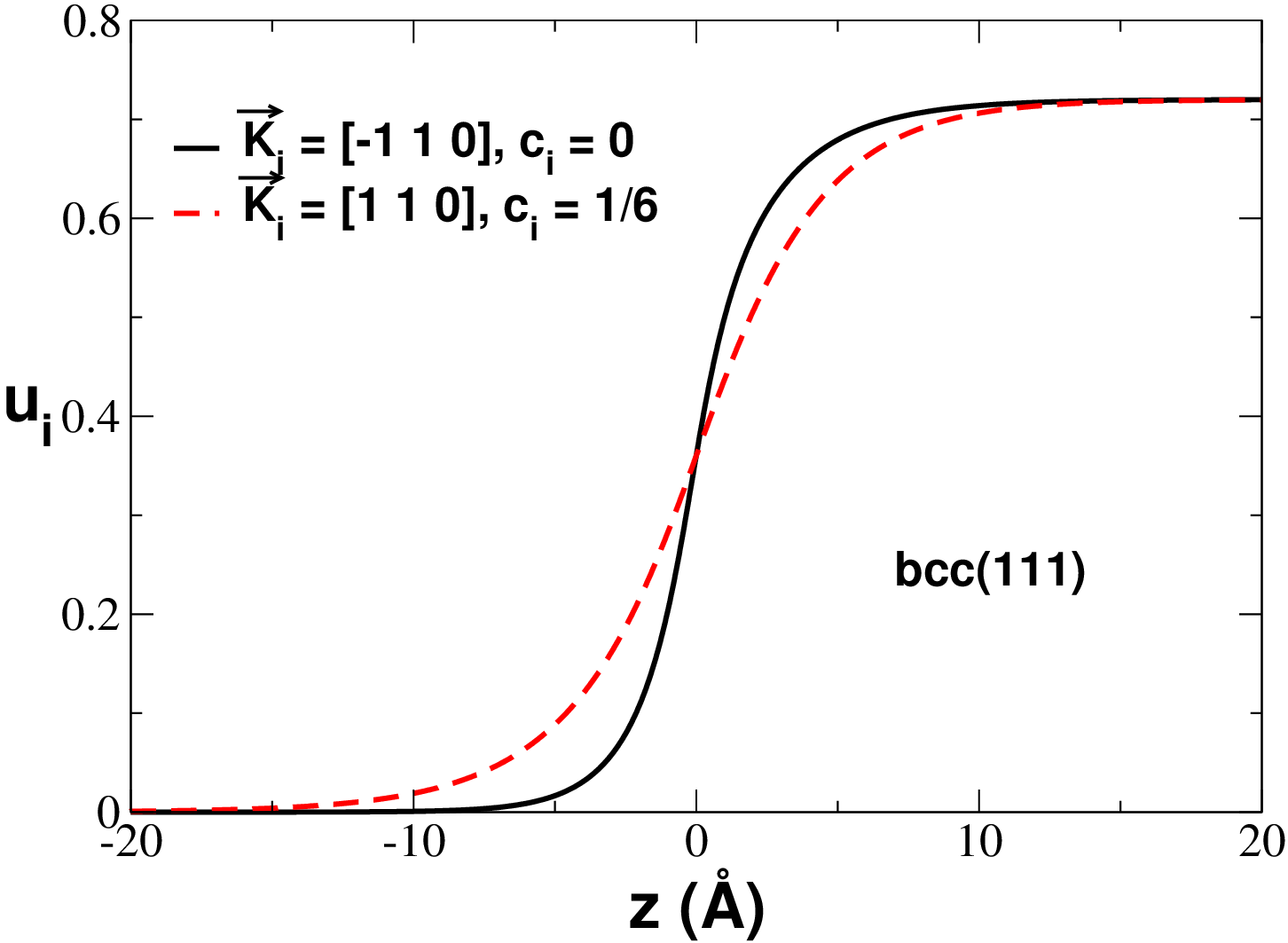} 
	\label{fig:1c}
	}
\caption{(Color online)  Equilibrium nonlinear density wave profiles across the solid-liquid interface obtained by GL theory with input parameters from MD simulations with the MH(SA)$^2$ potential for three crystal faces: \subref{fig:1a} (100), \subref{fig:1b} (110), and \subref{fig:1c} (111).}
\label{profiles}
\end{figure}

In this section, we compare kinetic coefficients predicted by the MC theory, the GL theory, and MD simulations with the MH(SA)$^2$ potential for Fe. The values of input parameters from MD simulations are listed in Table \ref{tabparam}. To compare the main result of the last section Eq. \eqref{finalresult} to the prediction of the MC theory \cite{MC1991,Chernov2004}, it is convenient to express $\mu$ in terms of the effective widths $\xi_{\vec{K}_i}$ of density wave $u_i$-profiles defined by 
\begin{equation}\label{width}
\xi_{\vec{K}_i}^{-1} = \int \, dz \left(\frac{du_i}{dz}\right)^2.
\end{equation}
and also introduce explicitly the correlation length of the liquid corresponding to the inverse half-width of the liquid structure factor, $\xi_b=\left( -S''(|\vec{K}_i|)/ 2 S(|\vec{K}_i|)\right)^{1/2}$. Using those definitions, Eq. \eqref{finalresult} can be rewritten in the form
\begin{eqnarray}\label{MCresult}
\mu=\frac{L}{k_BT^2_m}\frac{S(\vec{K}_i) \xi_b}{N_1\tau_L(\vec{K}_i) A_s},
\end{eqnarray}
where 
\begin{equation} \label{As}
A_s=\frac{1}{N_1}\sum_{\vec{K}_i} \xi_b / \xi_{\vec{K}_i}
\end{equation}
is a dimensionless anisotropy factor that depends on the orientation of the crystal face through the effective widths of density wave profiles; $N_1=12$ is the number of principal reciprocal lattice vectors for bcc lattices.
Remarkably, the expression for $\mu$ defined by Eq. \eqref{MCresult}, which has been derived here formally from GL theory, is identical to the one of the MC theory. A main difference, however, is that in GL theory, the $u_i$-profiles used to compute the widths defined by Eq. \eqref{width} and hence the anisotropy factor $A_s$ defined by Eq. \eqref{As} are nonlinear solutions of the equilibrium GL equation, e.g. Eq. \eqref{EOM3} for $V=\alpha=0$ for the (110) orientation. The different $u_i$ profiles across the solid-liquid interface are nonlinearly coupled through cubic and quartic terms in the free-energy density and need to be determined through a numerical solution of the equilibrium GL equations for the different set of $u_i$, with the set of $u_i$ depending on crystal orientation \cite{bcc1}, e.g. numerically solving Eq. \eqref{EOM3} for $u_0$, $v_0$ and $w_0$ for the (110) crystal face.  Numerically computed density wave profiles for the $(100)$, $(110)$, and $(111)$ crystal faces using input parameters from MD simulations with MH(SA)$^2$ potential \cite{bcc1} are plotted in Fig. \ref{profiles} ($\xi_b \sim 3.96 \AA$ for this potential). In contrast, in the calculation of kinetic anisotropy, MC estimate effective widths of density wave profiles using a truncated density functional theory derived in an earlier paper \cite{MC2}. The truncated density functional theory is a linear theory that predicts density waves profiles near the liquid and yields
\begin{equation} \label{widthMC1}
\xi_{\vec{K}_i}=\xi_b \left|\hat{K}_i \cdot \hat{n} \right|
\end{equation} 
for mixed transverse and longitudinal density waves with finite $\hat{K}_i \cdot \hat{n}$, and 
\begin{equation} \label{widthMC2}
\xi_{\vec{K}_i}=\left(\xi_b / |\vec{K}_i| \right)^{1/2} \equiv \xi_T
\end{equation}
for transverse density waves with $\hat{K}_i \cdot \hat{n}=0$. Then the dimensionless anisotropy factor can be approximated as
\begin{equation} \label{MCAs}
A^{MC}_s(\hat{n}) = \frac{1}{N_1} \left( \sum_T {\xi_b \over \xi_T} + \sum_{N.T.} {1\over \left|\hat{K}_i \cdot \hat{n} \right|} \right),
\end{equation}
where the summation is over transverse density waves and non-transverse density waves, respectively. The dimensionless anisotropy factor estimated by a linear theory exhibits the $\xi_b$ dependence through the transverse density waves. Thus the anisotropy in kinetic coefficient estimated by a linear theory is not universal but depends on the details of the interatomic potentials. In contrast, the full nonlinear density waves profiles are solved in GL theory, hence $\xi_{\vec{K}_i}$ can be evaluated directly using Eq. \eqref{width} without any approximations. It is convenient to express Eq. \eqref{width} in terms of the dimensionless length $\tilde{z}\equiv z/\xi_b$ and the rescaled amplitude $\tilde{u}_i \equiv u_i/u_s$, 
\begin{equation}
\xi_{\vec{K}_i}^{-1} \equiv \frac{u_s^2}{\xi_b} \,    c(\hat{K}_i; \hat{n}),
\end{equation}
where we define the dimensionless spatial integration of the derivative of density waves
\begin{equation}
c(\hat{K}_i; \hat{n}) \equiv \int \, d\tilde{z} \left(\frac{d\tilde{u}_i}{d \tilde{z}}\right)^2.
\end{equation}
The function $c(\hat{K}_i; \hat{n})$ depends only on the RLV and the interface normal. It can be seen from Eq. \eqref{EOM3} for $V=\alpha = 0$ that once we introduce above dimensionless length $\tilde{z}$ and rescaled amplitude $\tilde{u}$, these coupled Euler-Lagrange equations become independent of the liquid structure factor and give rise to universal nonlinear density wave profiles. Thus the function $c(\hat{K}_i; \hat{n})$ has a universal value regardless of the details of interatomic potentials. The universal values of $c(\hat{K}_i; \hat{n})$ are listed in Table \ref{ckn}. The dimensionless anisotropy factor computed by the GL theory is related to these universal values by 
\begin{equation} \label{MCAs}
A^{GL}_s(\hat{n}) = \frac{u_s^2}{N_1} \sum_i c(\hat{K}_i; \hat{n}).
\end{equation}
The GL theory predicts that the magnitude of $A_s$ depends on the solid amplitude square while the ratio of $A_s$ for different orientations remains the same.

 \begin{table}[b]
 \caption{
Values of $c(\hat{K}_i; \hat{n})$ and dimensionless anisotropy factors calculated using the MC theory and the GL theory.}
\centering
\begin{tabular*}{1.0\textwidth}%
     {@{\extracolsep{\fill}}c||cc |ccc|cc}  \hline
$\vec{n}$ & \multicolumn{2}{c|}{$(100)$}&\multicolumn{3}{c|}{$(110)$}&\multicolumn{2}{c}{$(111)$}
\\ \hline
$(\hat{K}_{i} \cdot \hat{n})^2$ & 0 &1/2& 0& 1/4 & 1 &0&2/3  \\ \hline
$c(\hat{K}_i ; \hat{n})$   & 0.37 & 0.28& 0.45 & 0.33&0.23&0.52&0.27    \\ \hline
$A^{MC}_s(\vec{n})$ & \multicolumn{2}{c|}{2.09}&\multicolumn{3}{c|}{2.07}&\multicolumn{2}{c}{2.33} \\ \hline
$A^{GL}_s(\vec{n})$ & \multicolumn{2}{c|}{0.161}&\multicolumn{3}{c|}{0.173}&\multicolumn{2}{c}{0.205} \\ \hline
\end{tabular*}
\label{ckn}
\end{table}

We compare in the first and the third column of Table \ref{kin_tab} the ratios of $\mu$ values for different crystal faces predicted by GL and MC theories. The $\mu$ values for GL theory are computed using Eq. \eqref{MCresult} with the widths $\xi_{\vec{K}_i}$ of density profiles (to evaluate $A_s$) computed using Eq. \eqref{width} and nonlinear equilibrium profiles shown in Fig. \ref{profiles} obtained from GL theory. The $\mu$ values for the MC theory are computed using the same Eq. \eqref{MCresult} but with the widths $\xi_{\vec{K}_i}$ predicted by Eqs. \eqref{widthMC1} and \eqref{widthMC2}. 
In addition, we list in the second column the ratios of $\mu$ calculated with GL theory using different ansatz for the weight of polygons that corresponds to the PFC free energy functional ($c_{ijk}=1/48$ and $c_{ijkl}=1/540$). To compare  the predictions of the two theories with results of MD simulations for the MH(SA)$^2$ EAM potential, we list in the fourth column of Table \ref{kin_tab} ratios of $\mu$ values computed using Eq. \eqref{MCresult} of the MC theory with widths $\xi_{\vec{K}_i}$ extracted from fits of MD-computed equilibrium density wave profiles to hyperbolic tangent functions of the normal coordinate $z$ \cite{Sunetal2004}. Finally, in the fifth column, we list the most accurate predictions to date of ratios of $\mu$ values extracted from nonequilibrium MD simulations for the same MH(SA)$^2$ EAM potential \cite{Gaoetal2010} (which improve the values previously reported in \cite{Sunetal2004}).

\begin{table}[t]
\caption{The anisotropy of the kinetic coefficient for bcc lattices predicted by the present GL theory that assumes equal weights of geometrically distinct polygons ($c_{ijk}=1/8$ and $c_{ijkl}=1/27$  \cite{shih, bcc1}) and the GL theory with normalization coefficients derived from the PFC model \cite{Wu07} that is equivalent to counting all repeats of those polygons ($c_{ijk}=1/48$ and $c_{ijkl}=1/540$), where both calculations use the full nonlinear equilibrium density wave profiles shown in Fig. (\ref{profiles}), the Mikheev-Chernov (MC) theory \cite{MC1991} with profile widths obtained from a linearized theory near the liquid (given by Eqs. \eqref{widthMC1} and \eqref{widthMC2}), the MC theory with widths of density wave profiles extracted from MD simulations \cite{Sunetal2004}, and by nonequilibrium MD simulations \cite{Gaoetal2010}.}
\centering
\begin{tabular*}{1.0\textwidth}%
     {@{\extracolsep{\fill}}cccccc}  \hline \hline
 & GL theory & GL theory & MC Theory & MC Theory  & MD \\  
 & (Coef. from Ref. \cite{shih, bcc1})   & (Coef. from Ref. \cite{Wu07})       & (Linear theory) &  (MD Profiles) &  \\ \hline
$\mu_{100}/\mu_{110}$ & 1.06 & 1.07 &  0.99 & 1.14 & 1.27 $\pm 0.11$  \\ 
$\mu_{100}/\mu_{111}$ & 1.27 & 1.39 & 1.12 & 1.23  & 1.26 $\pm 0.07$  \\ 
\end{tabular*}
\label{kin_tab}
\end{table}

The comparison of the first three columns and the fifth column in Table \ref{kin_tab} shows that the GL theory yields overall an improved prediction of the anisotropy of $\mu$. It better predicts the ratio $\mu_{100}/\mu_{111}$ and yields at least the correct ordering $\mu_{100}>\mu_{110}$ even if the ratio $\mu_{100}/\mu_{110}$ departs from the MD value (the ratio $\mu_{100}/\mu_{110}=1.06$ falls just at the lower end of the 95 percent confidence interval of the estimated MD value $1.27 \pm 0.11$ and has thus a relatively high probability of being lower than the true MD value). The comparison of the first and fourth columns indicates that a main contributing factor to this improvement is the fact that GL theory uses nonlinear density wave profiles with widths that better match the MD-calculated equilibrium profiles than the width predicted by Eqs. \eqref{widthMC1} and \eqref{widthMC2} used in the linear MC theory. 

In addition to the comparison of the anisotropy of kinetic coefficients, we compare the magnitude of kinetic coefficients predicted by the MC theory, the GL theory, and MD simulations. The kinetic coefficients are computed using Eq. \eqref{MCresult}, and the relaxation time of liquids measured from MD simulation is $0.57 \pm 0.05$ ps. The magnitude of $\mu$ predicted by the MC theory is an order of magnitude smaller than that measured from MD simulations, see Table \ref{mag_mu}. Underestimation for the magnitude of $\mu$ by the MC theory is shown in previous studies for Fe, Pb, Ni, and Lennard-Jones systems \cite{Sunetal2004, BGJ1982, MC1991, Pb, Ni}. In contrast, the magnitude of $\mu$ computed by GL theory is comparable with those found in MD simulations, since the dimensionless anisotropy factor $A_s$ computed by GL theory is obtained through the integration of spatial derivative of full nonlinear density waves profiles, see Table \ref{ckn}. 

Furthermore, the GL theory yields an analytical relation between two important interfacial quantities, namely the interfacial energy and the kinetic coefficients, as discussed below. Under the isotropic approximation, the interfacial energy derived from GL theory for bcc-liquid interfaces at equilibrium is proportional to the solid amplitude square \cite{shih, bcc1}, 
\begin{eqnarray} \label{gamma_iso}
\gamma_{iso} = \frac{n_0 k_B T_m u_s^2} {6} \sqrt{a_2b}.
\end{eqnarray}
The corresponding isotropic density wave profile is
\begin{eqnarray}
u = \frac{u_s}{2} \left(  1+ \tanh{\left( {\sqrt{3}z \over 2 \xi_b}\right)} \right),
\end{eqnarray}
which gives rise to $\xi_{\vec{K}_i}^{-1} =   \,  (\sqrt{3}/6) u_s^2 \xi_b^{-1}$ and the dimensionless anisotropy factor
\begin{eqnarray}\label{iso_A}
A_s  = \frac{\xi_b} {\xi_{\vec{K}_i}} =\frac {\sqrt{3}} {6} u_s^2.
\end{eqnarray}
Thus the magnitude of $\mu$ is proportional to the inverse of  $u_s^2$. Since both interfacial energy and kinetic coefficient are related to the solid amplitude square, we can relate these two quantities using Eq. \eqref{MCresult}, \eqref{gamma_iso}, and \eqref{iso_A},  
\begin{eqnarray}
\mu_{iso}=\frac{n_0 \, \xi_b^2 \, L }{3  \tau_L(\vec{K}_i) T_m} \frac{1}{\gamma_{iso}}.
\end{eqnarray}
The interfacial energy is inversely proportional to the kinetic coefficient, and these two interfacial quantities are  related through bulk liquid properties and latent heat in the GL theory.

\begin{table}[b]
\caption{The magnitude of the kinetic coefficient for bcc lattices predicted by the present GL theory that assumes equal weights of geometrically distinct polygons ($c_{ijk}=1/8$ and $c_{ijkl}=1/27$  \cite{shih, bcc1}), the Mikheev-Chernov (MC) theory \cite{MC1991} with profile widths obtained from a linearized theory near the liquid (given by Eqs. \eqref{widthMC1} and \eqref{widthMC2}), and by nonequilibrium MD simulations \cite{Gaoetal2010}. The unit of the kinetic coefficient is cm/(s$\cdot$ K).}
\centering
\begin{tabular*}{0.8\textwidth}%
     {@{\extracolsep{\fill}}cccc}  \hline \hline
 & GL theory  & MC Theory &   MD \\  
   & (Coef. from Ref. \cite{shih, bcc1})       & (Linear theory)  &  \\ \hline
$\mu_{100}$ & $64.68 \pm 5.67$ & $4.98 \pm 0.44$  &  $78.23 \pm 4.47$  \\ 
$\mu_{110}$ & $60.19 \pm 5.28$ & $5.03 \pm 0.44$ &  $61.67 \pm 4.11$  \\ 
$\mu_{111}$ & $50.80\pm 4.46$ & $4.47 \pm 0.39$ &  $62.08 \pm 2.26$  \
\end{tabular*}
\label{mag_mu}
\end{table}

\section{Concluding remarks and outlook}

The remaining discrepancy between MD simulations and GL theory is likely due to the over-simplified representation of the crystal density field in terms of the minimal set of principal RLV, which ignores contributions of higher order reciprocal lattice vectors. Interestingly, this representation yields a prediction of the anisotropy of $\mu$ in the GL theory that is independent of details of the interatomic potentials, which only enter in the theory through the amplitude of density waves in the solid $u_s$ and liquid structure factor properties. While those properties influence the magnitude of $\mu$, they do not influence its anisotropy because the shape of the density wave profiles are independent of $u_s$ and liquid structure factor properties up to a common multiplicative factor of the amplitude for all profiles and up to a common scaling factor of length for all widths, respectively. For the same reason, the anisotropy of the solid-liquid interfacial free-energy predicted by GL theory was found previously to be independent of details of interatomic potentials \cite{bcc1}. For a realistic crystal density field represented by a large set of RLV, the anisotropy of $\mu$ is expected to generally depend on the interatomic potential as found in several MD studies for different crystal structures \cite{Hoytetal1999,Hoytetal2002,Hoytetal2003,Sunetal2004,Gaoetal2010,Monketal2010,MoV}.
Thus extending GL theory to include more reciprocal lattice vectors could potentially give rise to a better prediction of kinetic anisotropy. 

In addition to the anisotropy, another interesting and unexplained aspect of MC theory is the magnitude of $\mu$.  In a previous MD study, Monk \emph{et al.} \cite{Monketal2010} proposed several techniques to correctly account for the temperature rise associated with latent release during free solidification MD simulations. The techniques were applied to an EAM model of fcc Ni and the authors found that the value of $\mu$ was approximately a factor of two larger than the $\mu$ computed without the temperature spike correction.  If we make the crude assumption that a similar factor of two can be applied to all previous MD studies (see the summary provided in Hoyt \emph{et al.} \cite{Hoytetal2003}), then it appears the MC model underestimates the kinetic coefficient in fcc crystals by a factor of roughly 3-4.  In this comparison various properties of the liquid, such as the structure function and the relaxation time, were estimated from the hard sphere system.  In the case of MD simulations of bcc Fe, Gao \emph{et al.} \cite{Gaoetal2010} have accounted for the effect of latent heat release and, as summarized in Table \ref{mag_mu}, the value of $\mu$ is an order of magnitude higher than the MC prediction.  Here again the kinetic coefficient is found to be about a factor of two higher than previous MD estimates for Fe \cite{Sunetal2004}.  Therefore it is safe to conclude that the MC model consistently underestimates the magnitude of the kinetic coefficient, the deviation is a factor of $\sim$ 3-4 for fcc and $\sim$ 10 for bcc.  It should be noted, however, that a preliminary MD study of the bcc elements \cite{MoV}, i.e. without an interface temperature correction, concluded that there is closer agreement with MC theory for the case of Mo and V than had been observed for Fe, which suggests that details of the interatomic potential not included in the MC treatment may be playing a role in bcc systems.  

To further elucidate the trend of kinetic coefficient with crystal structure and interatomic potential, a comparison of the GL model developed here to detailed MD simulations of other bcc, as well as fcc, systems is warranted. This comparison will require to extend the present calculation to other crystal structures. This should be possible by building on recent progress to reproduce quantitatively the anisotropy of the fcc-liquid interface with two different sets of density waves \cite{WuKarma2014}.  Such a comparison will also make it possible to explore more systematically the inverse relationship between the kinetic coefficient and the interfacial free-energy predicted by GL theory in this study.

\emph{Acknowledgments}:
During the initial stage of this work, the work of K.-A.W. and A.K. was supported by US DOE Award No. DE-FG02-92ER45471.
During the completion of this work, the work of C-H.W. and K-A.W. was supported by the National Science Council of Taiwan (NSC102-2112-M-007-007-MY3) and the support from National Center for Theoretical Sciences, Taiwan, and the work of A.K. was supported by US DOE Award No. DEFG02-
07ER46400. We also wish to thank Mark Asta for valuable discussions. 



\appendix

\end{document}